\documentclass[journal,onecolumn]{IEEEtran}
\usepackage{graphicx}
\usepackage{subfigure}
\usepackage{float}
\usepackage{amsmath}
\usepackage{epstopdf}
\usepackage{cases}
\usepackage{color}
\usepackage{stfloats}
\usepackage{algorithm}
\usepackage{algpseudocode}
\usepackage{amsmath}
\usepackage{ulem}

\ifCLASSINFOpdf

\else

\fi

\usepackage{caption}
\usepackage{mathtools}

\begin{document}

	\title{Integrated Sensing and Communication enabled Multiple Base Stations Cooperative Sensing Towards 6G}

	\author{
		Zhiqing Wei,~\IEEEmembership{Member,~IEEE,}
		Wangjun Jiang,~\IEEEmembership{Student Member,~IEEE,}\\
		Zhiyong Feng,~\IEEEmembership{Senior Member,~IEEE,}
		Huici Wu,~\IEEEmembership{Member,~IEEE,}
		Ning Zhang,~\IEEEmembership{Senior Member,~IEEE,}\\
		Kaifeng Han,~\IEEEmembership{Member,~IEEE,}
		Ruizhong Xu,~\IEEEmembership{Student Member,~IEEE,}
		Ping Zhang,~\IEEEmembership{Fellow,~IEEE}
		\\
		\thanks{Zhiqing Wei (corresponding author), Wangjun Jiang, Zhiyong Feng, Ruizhong Xu, and Ping Zhang are with the Key Laboratory of Universal Wireless Communications, Ministry of Education, School of Information and Communication Engineering, Beijing University of Posts and Telecommunications, Beijing 100876, China;

        Huici Wu is with the National Engineering Research Center of Mobile Network Technologies, Beijing University of Posts and Telecommunications, Beijing 100876, China;

        Ning Zhang is with the Department of Electrical and Computer Engineering, University of Windsor, Windsor, ON, N9B 3P4, Canada;

        Kaifeng Han is with China Academy of Information and Communications Technology, Beijing 100191, China.}}

	\maketitle

	\begin{abstract}
		Driven by the intelligent applications of sixth-generation (6G) mobile communication systems
		such as smart city and autonomous driving,
		which connect the physical and cyber space,
		the integrated sensing and communication (ISAC) brings a revolutionary change
		to the base stations (BSs) of 6G by integrating radar sensing and communication
		in the same hardware and wireless resource.
		However, with the requirements of long-range and accurate sensing in
		the applications of smart city and autonomous driving,
		the ISAC enabled single BS still has a limitation in the sensing range and accuracy.
		With the networked infrastructures of mobile communication systems,
		multi-BS cooperative sensing is a natural choice satisfying the
		requirement of long-range and accurate sensing.
		In this article,
		the framework of multi-BS cooperative sensing is proposed,
		breaking through the limitation of single-BS sensing.
		The enabling technologies, including unified ISAC performance metrics,
		ISAC signal design and optimization, interference management,
		cooperative sensing algorithms,
		are introduced in details. The performance evaluation results are provided to
		verify the effectiveness of multi-BS cooperative sensing schemes.
		With ISAC enabled multi-BS cooperative sensing (ISAC-MCS),
		the intelligent infrastructures connecting physical and cyber space
		can be established,
		ushering the era of 6G promoting the intelligence of everything.
	\end{abstract}


	\IEEEpeerreviewmaketitle

	\section{Introduction}\label{sec:introduction}

	\begin{figure*}[htbp]
		\includegraphics[scale=0.1]{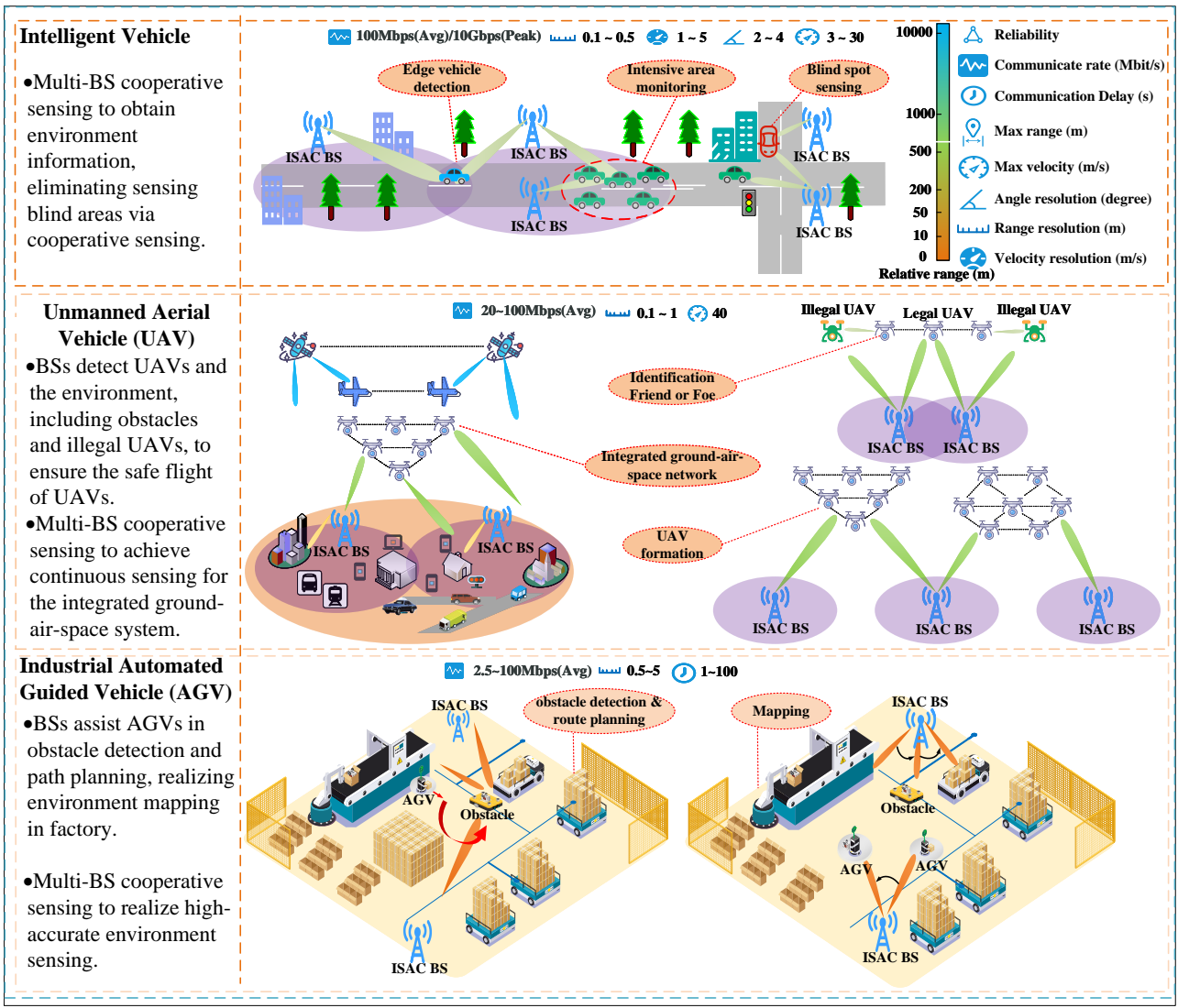}
		\centering
		\setlength{\abovecaptionskip}{2 mm}
		\caption{Three scenarios and eight use cases of ISAC enabled multiple BSs cooperative sensing. The required key parameter indicators are marked in the legends. The beam colors indicate the maximum ranges in the various use cases.}
		\label{fig:scenarios}
	\end{figure*}

	With the advancement of the sixth-generation (6G) mobile communication systems,
	the technologies of Internet of Things (IoT), artificial intelligence (AI),
	big data, and automation are reconstructing traditional industries \cite{6G_1},
	yielding the intelligent applications such as smart city and autonomous driving.
	These applications urgently need new information infrastructures
	with deep integration of sensing and communication \cite{6G_2}.
		As an important infrastructure supporting these emerging applications,
		the	mobile communication system is gradually evolving into a unified infrastructure
		with the integrated sensing and communication (ISAC) technology,
		which is one of the key potential technologies of 6G mobile communication systems.

		ISAC empowers current base stations (BSs) with sensing capabilities,
		allowing mobile communication systems to provide sensing services to civilians \cite{ISAC_1}.
		There are three levels of ISAC, namely device sharing, resource sharing, and waveform sharing.
		ISAC effectively reduces hardware cost and improves spectrum efficiency.
	The ISAC BS has the following advantages.

	\begin{itemize}
		\item \textbf{High resource utilization:}
		The sensing and communication functions reuse the same software,
		hardware and spectrum resources, improving resource utilization \cite{ISAC_1}.
		\item \textbf{Long-distance sensing:}
		The power of BS is high, owning excellent performance in long-distance sensing.
		\item \textbf{Mutual benefit between sensing and communication:}
		The sensing function assists communication in beamforming and
		beam alignment.
		Communication assists sensing in providing the prior information of targets.
	\end{itemize}

	However, single-BS still has shortcomings in sensing range and accuracy.
	Using the networked infrastructure of mobile communication systems,
	multi-BS cooperative sensing is promising to break through the limitation of single-BS sensing.
		Fig. \ref{fig:scenarios} shows three scenarios and eight use cases of
		multi-BS cooperative sensing \cite{ISAC_1},
		which includes intelligent vehicles,
		unmanned aerial vehicles (UAVs) and industrial automated guided vehicles (AGVs).
		In these scenarios with complex environment and high-dynamic targets,
		multi-BS cooperative sensing not only
		improves the successful probability of detection,
		but also improves the accuracy of parameter estimation and realizes continuous sensing
		of high-dynamic targets.
		Key performance indicators that meet the requirements
		for these use cases are listed in Fig. \ref{fig:scenarios}.
	Although multi-BS cooperative sensing has the advantages of extending sensing range,
	improving sensing accuracy, and improving sensing successful probability,
	there are the following challenges in realizing ISAC enabled multi-BS cooperative sensing (ISAC-MCS).

	\begin{itemize}
		\item \textbf{Difference in the performance metrics of sensing and communication:}
		The performance metrics of communication measure the transmission efficiency
		of random signals.
		While the {performance metrics of sensing measure the
			performance of detection and parameter estimation of targets}
		in the deterministic state of physical space.
		{The difference between them will bring difficulties in the
			design and optimization of ISAC signal.}
		{\item \textbf{Interference management in multi-BS cooperative sensing:}
			The interference in multi-BS cooperative sensing is complex, including the
			communication mutual interference, the sensing mutual interference, the mutual interference between communication and sensing, etc.}
		\item \textbf{Contradiction between the signals of sensing and communication:}
		There is a contradiction between the random communication signal and the structural sensing signal, which brings challenges to ISAC signal design and optimization.
		\item \textbf{Limitation of synchronization:}
		The synchronization in mobile communication systems is not designed for sensing.
		The cooperative sensing methods suitable to
		mobile communication systems need to be designed.
	\end{itemize}

	Facing the above challenges, the framework of ISAC-MCS is designed in this article.
		There are some related works about the ISAC enabled cooperative sensing,
		such as the perceptive mobile network (PMN) \cite{PMN},
		which originally proposed a framework of ISAC enabled mobile/cellular network.
		Then, a large number of target monitoring terminals
		are deployed in PMN to enhance the sensing capability
		\cite{WCM_PMN}.
		PMN focuses on the cooperative sensing among BS and user equipments (UEs).
		Compared with PMN, ISAC-MCS proposed in this article focuses on the framework of
		multi-BS cooperative sensing, including cooperative active sensing, cooperative passive sensing,
		cooperative active and passive sensing with
		data-level and signal-level sensing information fusion.
		Details of the framework of ISAC-MCS will be described in Section \ref{sec:Fra-3}.
	In addition, an in-depth analysis of enabling technologies supporting ISAC-MCS is provided, including unified ISAC performance metrics,
	ISAC signal design and optimization, interference management, and cooperative sensing algorithms.
	Finally, performance evaluations are provided
	to verify the effectiveness of ISAC-MCS.

	\section{Framework of ISAC enabled Multi-BS Cooperative Sensing}
	\label{sec:Fra}

	\begin{figure*}[ht]
		\includegraphics[scale=0.25]{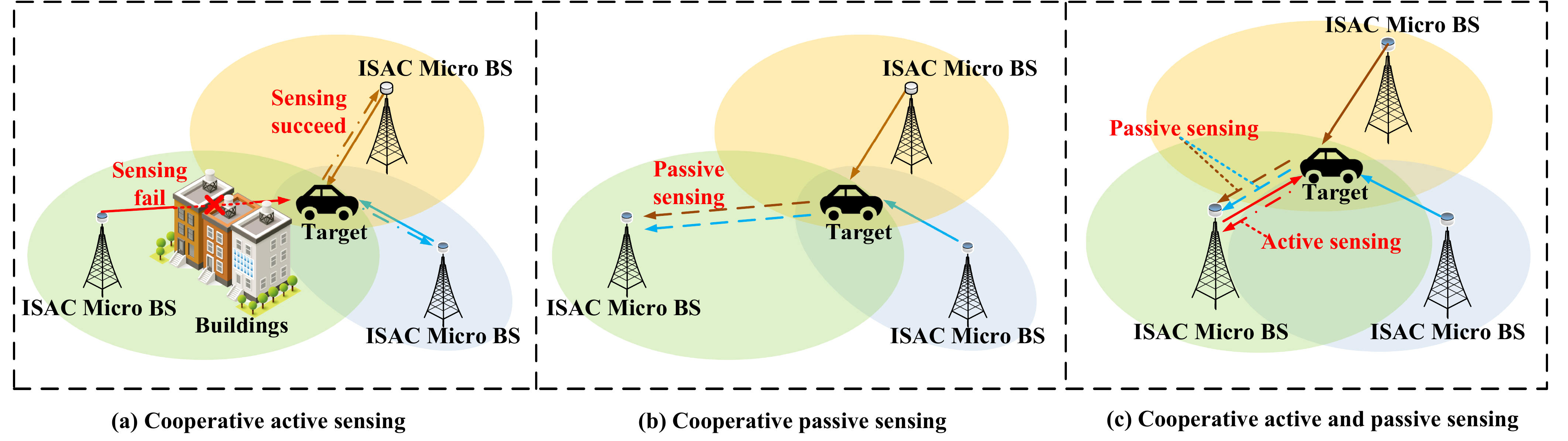}
		\centering
		\setlength{\abovecaptionskip}{2 mm}
		\caption{{Types of multi-BS cooperative sensing.}}
		\label{fig:2}
	\end{figure*}

	{
		According to the levels of sensing information fusion,
		multi-BS cooperative sensing methods can be classified into data-level fusion
		and signal-level fusion, which will be further introduced in Section \ref{sec:Fra-1}.
		According to the cooperation methods,
		multi-BS cooperative sensing methods can be classified into cooperative active sensing,
		cooperative passive sensing, cooperative active and passive sensing,
		which will be further introduced in Section \ref{sec:Fra-2}.
	}

	\subsection{Levels of Multi-BS Cooperative Sensing}\label{sec:Fra-1}
	{
		ISAC-MCS requires the fusion of sensing information
		from multiple BSs to achieve long-range and high-accurate sensing.
		Firstly, different BSs transmit their sensing information to the
		fusion center.
		Then, fusion center aligns the sensing information of multiple BSs
		in time and space dimensions.
		Finally, fusion center fuses the aligned sensing information for target sensing.
		The levels of sensing information fusion consist
		of data-level fusion and signal-level fusion.}

	\subsubsection{Data-level fusion}

	Data-level fusion is the fusion of sensing results from multiple BSs,
	which has low complexity.
	Data-level fusion only requires BSs to share the sensing results,
	which does not require high communication capacity.
	The space-time synchronization in data-level fusion
	is the unification of space coordinate systems and
	the alignment of sampling time of sensing results from multiple BSs.

	\subsubsection{Signal-level fusion}

	Signal-level fusion is the fusion of echo signals from multiple BSs,
	which has {higher complexity and higher accuracy} compared with data-level fusion.
	Signal-level fusion improves the signal-to-noise ratio (SNR) of echo signals by
	coherently fusing the echo signals from multiple BSs.
	Signal-level fusion requires high communication capacity and accurate space-time synchronization
	of multiple BSs.
	In terms of space synchronization,
	not only the coordinate systems of multiple BSs are required to be unified,
	but also the sensing areas corresponding to the echo signals of different BSs are required to be
	matched with each other.
	In terms of time synchronization, not only the sampling time needs to be aligned,
	but also the echo signals from multiple BSs need to realize clock synchronization
	to reduce the sensing error caused by clock offset.

	\subsection{Types of Multi-BS Cooperative Sensing}\label{sec:Fra-2}

		\subsubsection{Cooperative active sensing}\label{sec:cooper-1}
		As shown in Fig. \ref{fig:2}(a),
		all BSs send the echo signals to the fusion center for cooperative sensing.
		In cooperative active sensing,
		the unification of space coordinate systems and
		matching of sampling periods of the echo signals from multiple BSs are crucial
		when fusing the echo signals from multiple BSs.

		\subsubsection{Cooperative passive sensing}\label{sec:cooper-2}
		As illustrated in Fig. \ref{fig:2}(b),
		multiple BSs transmit ISAC signals to the targets and a BS
		receives the echo signals of the other BSs reflected by the targets and fuses the echo signals
		for target sensing,
		which is referred to
		multiple transmissions and single reception (MTSR).
		On the other hand, there are the schemes of single transmission and multiple receptions (STMR)
		and multiple transmissions and multiple receptions (MTMR), where
		there is a fusion center to fuse the echo signals of the receiving BSs for target sensing.
		In cooperative passive sensing,
		due to the separation of transmitters (TXs) and receivers (RXs),
		the space and time synchronization are crucial when fusing the echo signals from multiple
		receiving BSs.

		\subsubsection{Cooperative active and passive sensing}\label{sec:cooper-3}
		As illustrated in Fig. \ref{fig:2}(c),
		a BS receives its echo signal and the echo signals of other BSs.
		Then, these signals are fused to achieve high-accurate target sensing.
		Since all the signals are received by a BS,
		this type of cooperative sensing has a relatively small overhead on the communication
		interaction among BSs.

		In active sensing,
		since TX and RX are located in the same equipment,
		high-accurate synchronization between TX and RX can be realized.
		In passive sensing,
		the BS extracts the sensing information by receiving
		the echo signals of other BSs.
		Because of the separation of TX and RX in passive sensing,
		there are time offsets (TOs) and carrier frequency offsets (CFOs),
		which brings challenges for multi-BS cooperative sensing.

	\subsection{Framework of ISAC-MCS}\label{sec:Fra-3}

	\begin{figure*}[ht]
		\includegraphics[scale=0.45]{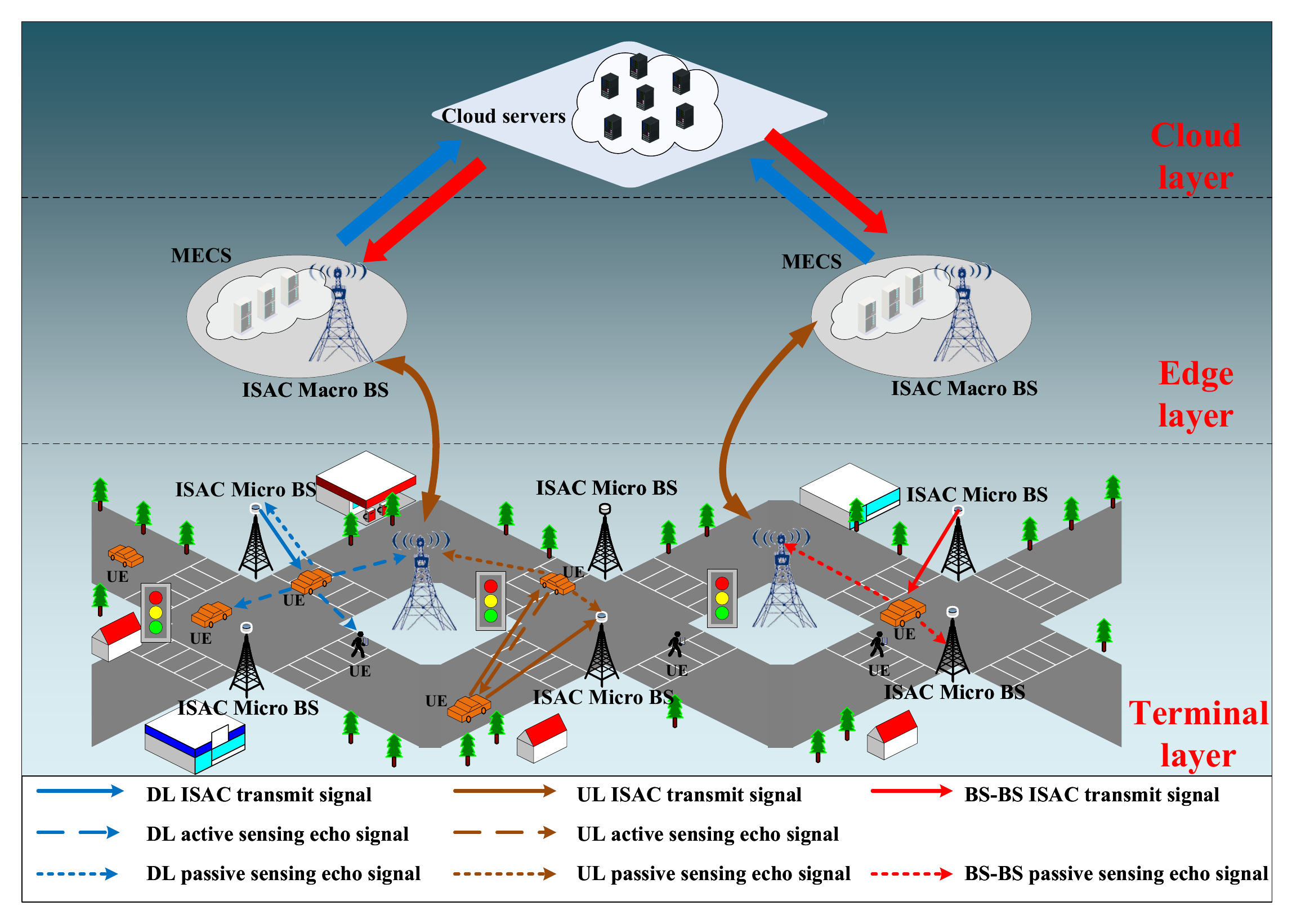}
		\centering
		\setlength{\abovecaptionskip}{0 mm}
		\caption{Framework of ISAC-MCS.}
		\label{fig:3}
	\end{figure*}

	Without loss of generality,
	intelligent vehicle network (IVN) \cite{WCM_PMN},
	which is regarded as the typical application scenario of ISAC,
	is taken as an example to reveal the framework of ISAC-MCS.
	As shown in Fig. \ref{fig:3},
	the functions of each layer are explained as follows.

	\subsubsection{Terminal layer}

	The terminal layer consists of micro BSs and UEs with ISAC capability.
	UEs have the advantage of flexible sensing.
	ISAC micro BS can be the fusion center in cooperative sensing.
	Cooperative sensing at the terminal level is classified into three categories.

	\begin{itemize}
		\item \textbf{Cooperative sensing among UEs:}
		UEs realize cooperative active sensing or cooperative active and passive sensing.
		Since UEs have high flexibility and mobility,
		cooperative sensing among UEs will overcome {the limitation of sensing coverage.}
		However, due to the limited wireless resources and computational capability of UEs,
		data-level fusion is the prior choice.
		\item \textbf{Cooperative sensing among UEs and micro BSs:}
		With a micro BS being the fusion center,
		the micro BS and UEs can perform cooperative active sensing.
		Besides, micro BS receives its echo signal and the uplink (UL)
		signals transmitted by UE.
		{Then, the cooperative active and passive sensing is realized,
			namely, the micro BS fuse the echo signal of the downlink (DL) signal
			transmitted by micro BS and the echo signal of the UL signal transmitted by UE.}
		This scheme has the advantages of flexible and long-range sensing.
		\item \textbf{Cooperative sensing among ISAC micro BSs:}
		Micro BSs can perform cooperative active sensing.
		Besides, a micro BS can receive {its echo signal and the echo signals} from
		other micro BSs to realize the cooperative active and passive sensing.
		This scheme has the advantages of accurate and long-range sensing.
	\end{itemize}

	{
		\subsubsection{Edge layer}
		The edge layer consists of ISAC macro BSs and {mobile edge computing servers (MECSs)}.
		ISAC macro BSs have a long sensing range.
		However, due to the blockage of obstacles,
		the blind areas exist,
		which can be eliminated via cooperative sensing.
		At the edge layer, cooperative sensing among macro and micro BSs can improve
		the sensing range and accuracy.
		The MECS at macro BS could act as the fusion center.
		Macro and micro BSs transmit ISAC signals on
		different frequency bands,
		which realizes multi-band sensing and obtains rich environmental sensing information,
		improving the sensing accuracy.

		\subsubsection{Cloud layer}
		The cloud layer has powerful computing and storage capabilities.
		The servers in the cloud layer aggregate the sensing information from
		edge layer and terminal layer, obtaining global sensing information.
		The cluster of cloud servers, acting as fusion center,
		aggregates and analyzes the global sensing information and
		performs decision making and task scheduling for vehicles.
		The cloud layer performs data-level cooperative sensing,
		extending the sensing range and realizing large-scale continuous sensing.

		\subsection{Effectiveness and Limitations of ISAC-MCS Framework}

		\subsubsection{Effectiveness of ISAC-MCS Framework}
		\begin{itemize}
			\item \textbf{Multi-layer cooperation:}
			In the framework of ISAC-MCS,
			terminal layer and edge layer can perform signal-level cooperative
			sensing to improve the sensing accuracy, in which the
			sensing results are aggregated into the cloud layer for data-level fusion to
			extend the sensing range.
			\item \textbf{Multi-node cooperation:}
			Multiple nodes including UEs, macro BSs and micro BSs
			could participate in cooperative sensing to
			improve the sensing accuracy and range
			with cooperative active sensing,
			cooperative passive sensing, or cooperative active and passive sensing.
		\end{itemize}

		\subsubsection{Limitation of ISAC-MCS Framework}
		ISAC-MCS requires relatively high synchronization accuracy among multiple BSs
		or the advanced sensing information fusion algorithms with
		synchronization error.
		Hence, ISAC-MCS will encounter the high computational complexity
		in sensing information fusion.
		Besides, facing the differentiated requirements of communication and sensing
		in various scenarios,
		appropriate time-frequency-space resources need to be allocated to multiple nodes
		including UEs, macro BSs and micro BSs,
		to coordinate the mutual interference between them,
		satisfying the requirements of sensing and communication.}

	\section{Key Enabling Technologies}\label{sec:Tec}

	In this section,
	the key enabling technologies for ISAC-MCS are introduced.
	Firstly, the unified performance metrics of communication and sensing are
	provided.
	Then, the design and optimization of ISAC signals
	are studied.
	Finally, the cooperative sensing algorithms
	suitable to mobile communication system are provided.

	\subsection{Unified Performance Metrics}\label{sec:Tec-1}

	The performance metrics of sensing and communication are the theoretical foundation
	of ISAC signal design and optimization.
    There are differences in the performance metrics
	of sensing and communication.
	It is necessary to find the unified performance metrics of them in ISAC system.

	Recently, communication mutual information (MI) and sensing MI
	in information theory are adopted
	to measure the performance of communication and sensing.
	The communication MI is defined as the MI between the transmitted signal and received signal.
	While the sensing MI is defined as the MI between the radar echo signal and
	radar sensing channel with the known transmitted signal.
	{
		In \cite{MI}, we derived the closed-form sensing and communication MIs
		in MIMO downlink system,
		which is further applied in ISAC signal optimization.}
	However, the existing research on sensing and communication MIs is
	still insufficient.
	For the multiple input multiple output - orthogonal frequency division multiplexing  (MIMO-OFDM) technology
	that is widely applied in mobile communication systems,
	the sensing MI of MIMO-OFDM ISAC system is rarely studied.
	Besides, the sensing MI in the scenario of ISAC-MCS
	is still vacant.

		The sensing and communication MIs in the scenario of ISAC-MCS need to be studied.
		The sensing MI
		is related to the characteristics of echo signals, interference, clutter, etc.
		To derive the sensing MI, the model of echo signals is first established.
		For multi-BS cooperative sensing,
		the fused signal of the echo signals from multiple BSs needs to be modeled.
		Then, the antenna arrays of multiple BSs are regarded as a virtual antenna array,
		and an equivalent radar sensing channel model is established
		as an equivalent MIMO-OFDM channel model
		according to the correlation between BSs.
		Finally, the sensing and communication MIs in the scenario of ISAC-MCS can be derived.
	Sensing and communication MIs yield the influence of
	the parameters including sequence structure,
	bandwidth, etc., on the performance of sensing and communication,
	which provide guideline for ISAC signal design and optimization.

		\subsection{Interference Management}\label{sec:Tec-4}

		The interference in ISAC-MCS system is more complicated
		than the interference in the scenario of single-BS sensing,
		which needs to be addressed in ISAC-MCS system.

		\subsubsection{Types of Interference}
		There are mainly two types of interference in ISAC-MCS system.
		\begin{itemize}
			\item \textbf{The communication mutual interference between multiple BSs:}
			When multiple BSs provide communication services to the UEs in the same area,
			the UEs will receive multiple downlink signals from different BSs,
			thus causing the communication mutual interference.

			\item \textbf{The mutual interference between communication and sensing:}
			When a BS receives the echo signal reflected by the target,
			it also receives the uplink communication signal sent by UE,
			thus causing the mutual interference between communication and sensing.
		\end{itemize}

		\subsubsection{Interference Management}
		Due to the complex mutual interference in the scenario of ISAC-MCS,
		the collaborative precoding of multiple BSs in space-time-frequency domains
		needs to be studied,
		taking into account the communication mutual interference, sensing mutual interference,
		and the mutual interference between sensing and communication.
		The optimal precoding design needs to
		be obtained to achieve appropriate tradeoff
		between sensing and communication.

	\subsection{ISAC Signal Design and Optimization}\label{sec:Tec-2}

	ISAC signal design and optimization are essential to improve
	the performance of sensing and communication in various scenarios.

	\subsubsection{ISAC signal design}

	Towards 6G, ISAC signals are designed based on the signals of mobile communication systems,
	which include single-carrier signals and multi-carrier signals.

	\begin{enumerate}
		\item \textbf{Single-carrier signals:}
		The single-carrier signals have the advantages of small power amplifier backoff,
		low peak-to-average power ratio (PAPR),
		and low energy,
		which include single carrier frequency domain equalization (SC-FDE),
		and discrete Fourier transform spreading OFDM (DFT-s-OFDM), etc.

		\item \textbf{Multi-carrier signals:}
		The spectrum efficiency and the PAPR of multi-carrier signals are high.
		Hence, it is necessary to reduce the PAPR of multi-carrier ISAC signals.
		Multi-carrier signals mainly include orthogonal time frequency space (OTFS),
		OFDM with cyclic prefix (CP-OFDM),
		filter bank multi-carrier (FBMC), etc.
	\end{enumerate}

In multi-BS cooperative sensing,
		the ISAC signal and frame structure need to be designed cooperatively
		to ensure the efficient fusion of sensing information from multiple BSs.
		The space-time-frequency resources of the ISAC signals of multiple BSs could be
		jointly optimized to coordinate the interference in ISAC-MCS system
		and improve the performance of communication and sensing.
		Take cooperative active sensing as an example,
		the downlink subframes of multiple BSs realizing sensing function
		needs to be aligned to improve the sensing efficiency.

	\subsubsection{ISAC Signal Optimization}

	With ISAC signal design,
	it is necessary to further optimize the ISAC signal according to
	the requirements of sensing and communication in various scenarios.
	Using a multi-objective optimization model,
	the sensing and communication MIs can be chosen as the objective function.
	The PAPR, transmit power, etc., can be selected as the constraints.
	The decision variables are the signal parameters in the space-time-frequency dimensions.

	Taking the signal optimization of MIMO-OFDM ISAC-MCS system
	as an example, the weighted sum of sensing and communication MIs is selected as the
	objective function, with the weight flexibly adjusting according to
	the requirement of sensing and communication.
	The joint precoding matrix of multi-BS in the space dimension,
	the subcarrier allocation, power allocation, and pilot design
	in time-frequency dimensions are jointly optimized.
	In addition, the constraints such as transmit power and PAPR are considered.

	\subsection{Cooperative Sensing Algorithms}\label{sec:Tec-3}

	Traditional multi-radar cooperative sensing algorithms have extremely
	high requirements on the synchronization accuracy and
	the deployment locations of radars,
	which cannot be directly applied in ISAC-MCS.
	Synchronization and signal processing algorithms for
	mobile communication systems are not initially designed for radar sensing.
	Therefore, it is necessary to design the
	multi-BS cooperative sensing algorithms,
	which fuse the sensing information from multiple BSs
	to improve the performance of sensing,
	including space-time registration algorithms and cooperative sensing algorithms.

	\subsubsection{Space Registration}

	In ISAC-MCS,
	space registration is a prerequisite for the efficient fusion of sensing information.
	The differences in geographic locations of different BSs lead to spatially
	unsynchronized sensing information,
	which degrades the performance of sensing information fusion \cite{SP_2}.

	For data-level fusion,
	space registration is to realize the unification of coordinate systems
	of multiple BSs and compensate for the systematic errors of sensing information fusion.
	For signal-level fusion,
	space registration realizes the matching of sensing areas of multiple BSs.
	To solve this problem,
	we propose an adjustable beam enabled space registration algorithm (AB-SRA)
	based on beamwidth adjustable beamforming algorithm (BABA) \cite{SP},
	which flexibly adjusts the side length of the sensing area of
	different BSs to achieve the matching of sensing areas
	by generating the sensing beam with adjustable beamwidth.

	\subsubsection{Time Registration}

	Similar to space registration,
	time registration is a prerequisite for efficient sensing information fusion.
	Deviations in clocks and sampling periods of multiple BSs
	lead to TOs of the sensing information of multiple BSs,
	which results in errors in sensing information fusion
	and degrades the performance of ISAC-MCS.

	For data-level fusion,
	time registration is
	to align the sensing information from different BSs in sampling time by filtering and
	interpolation.
	For signal-level fusion,
	time registration is to
	mitigate TOs of the sensing information from different BSs.
	For this purpose,
	one solution is to achieve clock synchronization of different BSs.
	However, synchronization algorithms for mobile communication systems
	hardly satisfy the requirement of signal-level cooperative sensing.
	Thus, the alternative solution is to fuse sensing information
	directly with the synchronization level of mobile communication system
	and compensate for the sensing error
	caused by TOs in the cooperative sensing algorithm.

	\subsubsection{Signal processing in cooperative sensing}

	In terms of the signal processing for ISAC-MCS, three types of cooperative sensing need to be addressed,
		including cooperative active sensing,
		cooperative passive sensing,
		and cooperative active sensing and passive sensing.
	Implementing ISAC-MCS with the
	synchronization level of mobile communication systems is the main challenge.

    \begin{itemize}
        \item \textbf{Cooperative active sensing and passive sensing:}
        In terms of cooperative active sensing and passive sensing, the transceivers of passive sensing are separated from each other
        in contrast to active sensing,
        which leads to the problem of time-frequency offsets.
        Actually, this problem can be eliminated in
        cooperative active and passive sensing.
        We propose the cross-correlation cooperative sensing (CSCC) based algorithm,
        which extracts the deviations of clock and carrier frequency
        in passive sensing by correlating the echo signals reflected from the target
        in active and passive sensing.
        Compared with cross-antenna cross-correlation (CACC) algorithm
        proposed in \cite{CACC},
        CSCC does not require the existence of line-of-sight (LOS)
        propagation paths between transceivers because
        it applies the echo signal of active sensing.

        \item \textbf{Cooperative active sensing:}
        In terms of cooperative active sensing,
        the data-level fusion has low complexity and relatively poor accuracy
        in parameter estimation
        due to the loss of sensing information in signal processing.
        On the contrary,
        signal-level fusion retains the original sensing information,
        thus enabling more accurate parameter estimation compared with data-level fusion.
        However, the complexity of signal-level fusion is high.

        Combining the advantages of data-level fusion and signal-level fusion,
        this article proposes a cooperative active sensing algorithm.
        Firstly, the data-level fusion algorithms \cite{KF}
        are applied
        to obtain the rough range and velocity of targets, generating
        the confidence region where the target is located
        and the confidence interval of the velocity.
        The signal-level cooperative sensing algorithm
        is iteratively applied to search for the accurate location of targets from the confidence region
        and search for the accurate velocity of targets from the confidence interval.
        With the confidence region discretized into smaller regions,
            the symbol matrices of multiple BSs are fused
            to obtain the smaller confidence region for the
            location estimation of target.
            The above operations run iteratively to generate
            accurate location estimation of target.
            Similarly, the velocity of target can also be estimated iteratively.

        \item \textbf{Cooperative passive sensing:}
        In terms of cooperative passive sensing,
            TOs and CFOs between TX and RX need to be eliminated.
            If the ideal line-of-sight (LoS) links between TX and RX exist,
            the traditional cross-correlation algorithms could be applied to eliminated TOs and CFOs
            between TX and RX.
            However, the most challenging research is the elimination of TOs and CFOs between
            TX and RX without LoS links using the cross-correlation algorithms.
            Then, the distance, velocity and angle of the targets could be estimated.

    \end{itemize}

	\section{Performance Evaluation}\label{sec:Per}
		In this section, the performance evaluation results of some enabling technologies are provided to verify the effectiveness of ISAC-MCS, including space registration, cooperative active sensing, and cooperative active and passive sensing.

			\subsection{Space Registration}\label{sec:Per-1}

			We consider the signal-level fusion, where multiple BSs detect the target from different directions. Therefore, the echo signals of different BSs are spatially asynchronous.
			As described in Section \ref{sec:Tec-3}, we propose AB-SRA with BABA, flexibly adjusting the sensing area of multiple BSs by generating sensing beams with adjustable beamwidth,
			realizing the matching of sensing areas of multiple BSs.

\begin{figure}[ht]
				\includegraphics[scale=0.8]{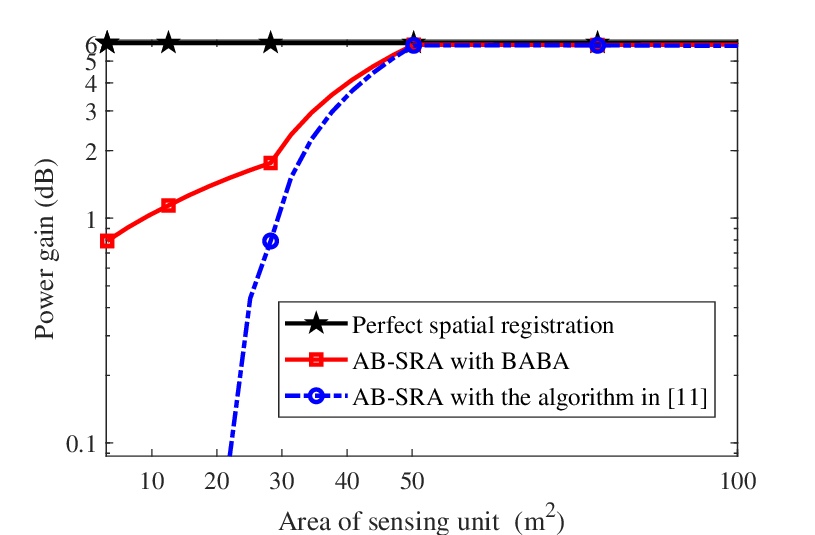}
				\centering
				\setlength{\abovecaptionskip}{0mm}
				\caption{The sensing performance improvement with space registration.}
				\label{fig:5}
			\end{figure}

			In the simulation, the number of BSs is set as 4, the size of the circular center antenna array is set as $32 \times 32$, the frequency and bandwidth of ISAC signals are set as 24 GHz and 100 MHz.
			As shown in Fig. \ref{fig:5}, under the perfect space registration, echo signals of four BSs can obtain a power gain of 6 dB through signal level fusion. When the area size of the sensing unit is larger than 30 $\rm{m}^2$,
			BABA can generate sensing beams of desired width,
			and the power gain of the fused echo signal is close to the performance of
			perfect space registration.
			Moreover, AB-SRA with BABA has better overall performance than AB-SRA
			with the typical beamforming algorithm \cite{BF}.

			\begin{figure}[ht]
				\includegraphics[scale=0.8]{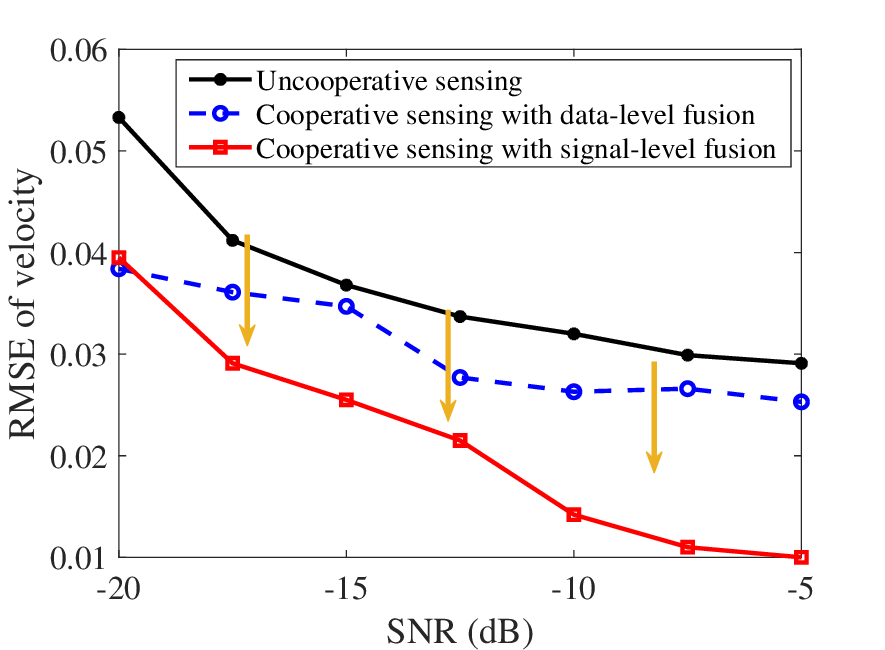}
				\centering
				\setlength{\abovecaptionskip}{0mm}
				\caption{Sensing performance improvement with cooperative sensing.}
				\label{fig:7}
			\end{figure}

			\begin{figure}[ht]
				\includegraphics[scale=0.8]{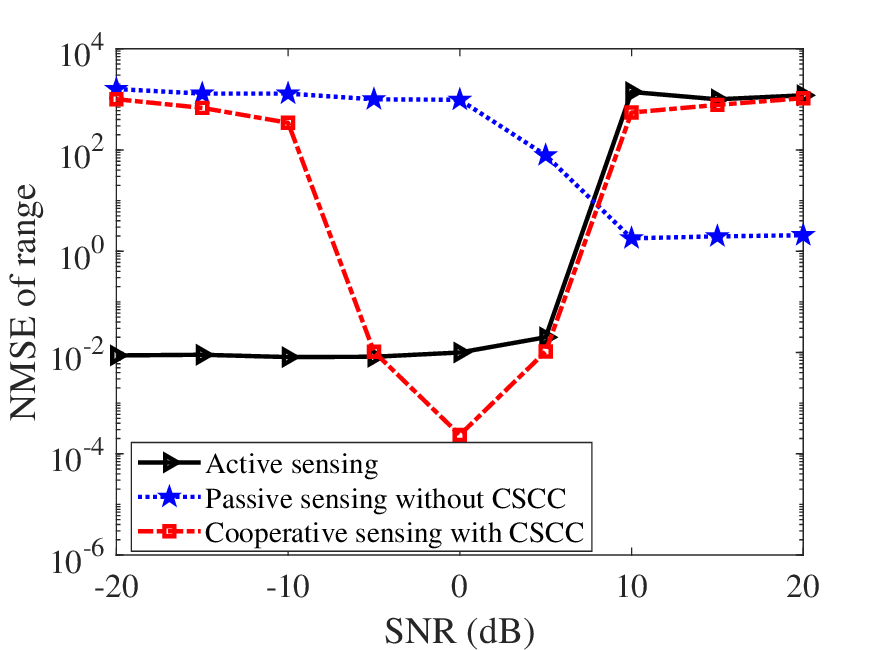}
				\centering
				\setlength{\abovecaptionskip}{0mm}
				\caption{Cooperative active and passive sensing.}
				\label{fig:6}
			\end{figure}

			\subsection{Cooperative Active Sensing}\label{sec:Per-3}

			With the synchronization level of mobile communication systems,
			the cooperative active sensing scheme is proposed in Section \ref{sec:Tec-3},
			reducing the computational complexity and the requirement of time synchronization.
			Fig. \ref{fig:7} shows the sensing performance achieved by cooperative active sensing and uncooperative sensing.
			The OFDM signal is used as the ISAC signal to detect targets.
			The frequency of the ISAC signal is set as 24 GHz, the frequency bandwidth of the ISAC signal is set as 93.1 MHz.
			The frequency and bandwidth of ISAC signals are set as 24 GHz and 93.1 MHz, respectively.
			The distance between the target and BS is set as 500 m, the velocity of the target is set as 27 m/s.
			As shown in Fig. \ref{fig:7}, the performance of signal-level fusion is better than that of
			data-level fusion with the same SNR,
			where data-level fusion implies a weighted average of the sensing results
			from multiple BSs.
			With the increase of SNR,
			the root mean square error (RMSE) of signal level fusion algorithm is decreasing
			faster than that of the data-level fusion algorithm.

		\subsection{Cooperative Active and Passive Sensing}\label{sec:Per-2}
		In terms of passive sensing,
		there are TOs and CFOs
		due to the separation of the TX and RX,
		which will reduce the sensing accuracy.
		Coincidently,
		this problem can be solved via the cooperation between active
		and passive sensing, as explained in Section \ref{sec:Tec-3}.
		The OFDM signal is used as ISAC signal to detect the target.
		The frequency and bandwidth of ISAC signals are set as 4 GHz and 123 MHz, respectively. The distance between the target and BS is set as 100 m.
		The SNRs of the echo signals of active sensing and passive sensing
		are set as 0 dB and from -20 dB to 20 dB, respectively.
        Fig. \ref{fig:6} shows the sensing performance achieved by active sensing, passive sensing and cooperative sensing.
        The ranging normalized MSE (NMSE) of cooperative sensing first decreases and then increases with the increase of the SNR of the passive echo signal, and reaches the lowest point when the SNR is 0 dB, and the target ranging accuracy reaches the optimum, which is due to the fact that the passive echo signal is similar to the active echo signal when the passive echo signal is 0 dB, and optimal results can be obtained when adopting the CSCC algorithm.
        Moreover, when the SNR of the passive echo signal is 0 dB, the red line is much lower than the black and blue line, which confirms that cooperative sensing with CSCC can greatly improve the performance of target sensing.

		We can conclude that, when the power of the echo signal of active sensing is
		much higher than that of passive sensing,
		the sensing performance of active sensing is the best.
        When the power of the echo signal of active sensing
		is much lower than that of passive sensing,
		the sensing performance of passive sensing is the best.
        However, when the received signal power of active sensing
		is close to that of passive sensing,
		the sensing performance of cooperative active and passive sensing with CSCC
		is the best.
		Overall, when the received signal power of active sensing
		is close to that of passive sensing, the sensing performance of cooperative active and passive sensing with CSCC is higher than that of uncooperative sensing.

	\section{Future Research Directions}\label{sec:Fut}

In this section, we discuss open problems and research
		opportunities in ISAC-MCS.

	\subsection{Unified ISAC Performance Metrics}\label{sec:Fut-1}

There are the following open problems
		for the unified ISAC performance metrics.

	\begin{itemize}
		\item \textbf{Complete ISAC Performance Metric System:}
		In order to design and optimize ISAC signals in various scenarios,
		it is necessary to establish a complete ISAC performance metric system.
		Sensing and communication MIs,
		as well as the relation between MI and other performance metrics
		such as detection probability, false alarm probability, minimum mean square error (MMSE), Cramer-rao lower bound (CRLB), etc., need further research \cite{MI_new}.
		\item \textbf{Sensing and Communication MIs with New Air-interface:}
		New air-interface technologies are proposed for 6G,
		such as orthogonal time frequency space (OTFS), holographic MIMO.
		The sensing and communication MIs with new air-interface technologies
		need to be studied.
	\end{itemize}

	\subsection{ISAC Signal Design and Optimization}\label{sec:Fut-2}

Although ISAC signal design and optimization are studied widely,
		there are still the following challenges.

	\begin{itemize}

		\item \textbf{Multi-dimensional ISAC signal:}
		ISAC signal is separately studied in space dimension or
		time-frequency dimensions.
		ISAC signal design and optimization in space-time-frequency multi-dimension need to be studied.

		\item \textbf{ISAC signal for ISAC-MCS:}
		ISAC signals in the scenario of
		single-BS sensing are widely studied.
		The ISAC signal design and optimization for ISAC-MCS
		are still in the infancy stage.

	\end{itemize}

	\subsection{Cooperative Sensing Algorithm}\label{sec:Fut-3}

	It is noted that the study on multi-BS cooperative sensing algorithms is still in
	the early stage.
	There are still challenges for the cooperative sensing algorithms.

	\begin{itemize}
		\item \textbf{Space-time registration:}
		The design of space-time synchronization
			supporting signal-level fusion
			is still challenging.

		\item \textbf{Sensing information fusion:}
		Sensing information fusion suitable to the synchronization level of
		mobile communication systems remains to be a challenge.

		\item \textbf{Multi-target sensing:}
		Multi-target sensing requires the matching of different targets.
		It is challenging to design low-complexity multi-target sensing algorithms.
	\end{itemize}

    \subsection{Attack and Defense in ISAC-MCS}\label{sec:Fut-4}
     Compared with traditional communication system, attack and defense in ISAC-MCS are more complex.
    There are some challenges for the attack and defense in ISAC-MCS.

    \begin{itemize}
        \item \textbf{Virtual node attack and defense in ISAC-MCS:}
        In multi-node cooperative communication systems,
        it is possible to disrupt the entire communication network
        by generating virtual malicious communication nodes.
        However, in ISAC-MCS, ISAC BSs can not only obtain the digital information of
        the nodes by means of communication, but also obtain
        the physical information of the nodes, such as location and velocity,
        by radar sensing, and then distinguish the real nodes from
        the virtual nodes through the coupling of the digital and physical information.
        Therefore, applying ISAC to achieve the coupling of digital
        and physical information
        to solve the attack and defense problem in ISAC-MCS is promising and still challenging.

        \item \textbf{Electromagnetic interference (EMI) attack and defense in ISAC-MCS:}
        EMI attack is intractable problem in both communication system and ISAC-MCS system. Moreover, compared with the communication system, the interference in ISAC-MCS is more complex. Therefore, it is difficult to address the EMI defense schemes in ISAC-MCS.

    \end{itemize}

	\section{Conclusion}\label{sec:Con}

    ISAC is one of the key 6G technologies that can revolutionize intelligent applications including smart city and autonomous driving.
    However, the ISAC enabled single BS still has a limitation in the sensing range and accuracy to meet the requirements of long-range and accurate sensing in the applications of smart city and autonomous driving.
    To address this problem, this paper proposes ISAC-MCS, which breaks through the limitation of single-BS sensing.
    ISAC-MCS has the advantages of long-range and high-accurate sensing, which can be achieved at a relatively low cost with the networked infrastructures of mobile communication systems.
    In this article, the framework of ISAC-MCS,
    as well as the enabling technologies including unified ISAC performance metrics,
    ISAC signal design and optimization, cooperative sensing algorithms,
    are introduced in details.
    The simulation results are shown to verify the effectiveness of ISAC-MCS.
    ISAC-MCS improves the sensing performance,
    which will promote the development of 6G applications and usher the era of 6G
    promoting the intelligence of everything.
    Finally, the future trends of ISAC-MCS are revealed.

    \normalem 
	\bibliographystyle{IEEEtran}
	\bibliography{reference}

\vfill

\end{document}